\documentclass{edp-jp4}
\usepackage{graphicx}
\usepackage{amssymb}
\usepackage{amsmath}

\begin{document}

\title{Random patterns generated by random permutations of natural numbers}
\author{G.Oshanin}\address{Physique Th\'eorique de
la Mati\`ere Condens\'ee (UMR 7600), Universit\'e Pierre et Marie
Curie - Paris 6, 4 place Jussieu, 75252 Paris
France}\secondaddress{Department of Inhomogeneous Condensed Matter
Theory, Max-Planck-Institute f\"ur Metallforschung,
Heisenbergstrasse 3, D-70569 Stuttgart, Germany}
\author{R.Voituriez}\sameaddress{1}
\author{S.Nechaev}\address{LPTMS, Universit\'e Paris Sud,
91405 Orsay Cedex, France }
\author{O.Vasilyev}\sameaddress{2}
\author{F. Hivert}\address{LITIS/LIFAR,
Universit\'e de Rouen,  76801 Saint Etienne du Rouvray, France}

\maketitle

\begin{abstract} We survey recent results on some one- and
two-dimensional patterns generated by random permutations of natural
numbers. In the first part, we discuss properties of random walks,
evolving on a one-dimensional regular lattice in discrete time $n$,
whose moves to the right or to the left are induced by the
rise-and-descent sequence associated with a given random
permutation. We determine exactly the probability of finding the
trajectory  of such a permutation-generated random walk at site $X$
at time $n$,  obtain the probability measure of different excursions
and define the asymptotic distribution of the number of "U-turns" of
the trajectories - permutation "peaks" and "through". In the second
part, we focus on some statistical properties of surfaces obtained
by randomly placing natural numbers $1,2,3, \ldots,L$ on sites of a
1d or 2d square lattices containing $L$ sites. We calculate the
distribution function of the number of local "peaks" - sites the
number at which is larger than the numbers appearing at
nearest-neighboring sites - and discuss some surprising collective
behavior emerging in this model.
\end{abstract}

\section{Introduction}

Properties of random  or patterns-avoiding permutations of series of
natural numbers  have been analyzed by mathematicians  for many
years. Studies of several problems emerging within this context,
such as, e.g., the celebrated Ulam's longest increasing subsequence
problem (see, e.g., Refs.\cite{baik,wilf,od} and references
therein), provided an entry to a rich and diverse circle of
mathematical ideas \cite{dia}, and were also found relevant to
certain physical processes, including random surface growth
\cite{sep,pra,jon,nec} or 2D quantum gravity (see, e.g.,
Ref.\cite{od}).

In this paper we focus on random permutations from a different
viewpoint addressing the following question: what are  statistical
properties of patterns, e.g., random walks or surfaces, when random
permutations are used as their generator? We note that such a
generator is different of those conventionally used, since here a
finite amount of numbers is being shuffled and moreover, neither of
any two numbers in each permutation may be equal to each other; this
incurs, of course, some correlations in the produced sequences of
random numbers.

In the first part of this paper, we consider a simple model of a
permutation-generated random walk (PGRW), first proposed and solved
in Ref.\cite{gr}. In this model, random walk evolves in discrete
time on a one-dimensional lattice of integers, and the moves of the
walker to the right or to the left are prescribed by the
rise-and-descent sequence characterizing each given permutation $\pi
= \{\pi_1, \pi_2, \pi_3, \ldots , \pi_l, \ldots , \pi_{n+1}\}$. In a
standard notation, the "rises" (the "descents") of the permutation
$\pi$ are such values of $l$ for which $\pi_l < \pi_{l+1}$ ($\pi_l >
\pi_{l+1}$).

We determine exactly several characteristic properties of such a
random walk, including  the probability ${\cal P}_n(X)$ of finding
the end-point $X_n$ of the PGRW trajectory at site $X$, its moments,
the probability measure of different excursions and the asymptotic
distribution of the number of the "U-turns" of the trajectories,
which, in the permutation language, corresponds to the number of
"peaks" and "through" in a given permutations.

In the second part, we focus on some statistical properties of
surfaces created by randomly distributing numbers $1,2,3, \ldots,L$
on sites of one- or two-dimensional lattices  containing $L$ sites.
A number appearing at the site $j$ determines the local height of
the surface. Denoting as local surface "peaks" such sites the number
at which exceeds the numbers appearing at neighboring sites, we aim
to calculate the probability $P(M,L)$ of having  $M$ peaks on a
lattice containing $L$ sites.

 In one dimension this can be done
exactly \cite{hiv}. We are also able to calculate the "correlation
function" $p(l)$ defining the conditional probability that two peaks
are separated by the interval $l$ under the condition that the
interval $l$ does not contain other peaks. In 2D, determining
exactly first three cumulants of $P(M,L)$, we define its asymptotic
form  using expansion in the Edgeworth series \cite{edg} and show
that it converges to a normal distribution as $L \to \infty$
\cite{hiv}. For 2D model, we will also discuss some surprising
cooperative behavior of peaks.

\section{Permutation-generated random walk}

Suppose there are two players - $A$ and $B$ going to play the
following mindless card game. They first agree on the value of each
card linearly ordering them, e.g., by putting suits  in the
bridge-bidding order: Clubs $<$ Diamonds $<$ Hearts $<$ Spades, and
adopting a convention that in a series of cards with the same suit
"two" is less than "three", "three" is less than "four" and etc,
while the ace has the largest value. So, in such a way, a deck of
cards is labeled $1,2,3, \ldots,52$. Then, the deck is shuffled, the
upper card is turned its face up, and our players start the game:
the second card is turned face up; if its value is higher than the
value of the first card, the player $A$ receives some unit of money
from the player $B$; if, on contrary, its value is less than the
value of the first card, player $B$ receives a unit of money from
the player $A$. At the next step, the third card is turned its face
up and its value is compared against the value of the second; if,
again, its value is higher than the value of the second card, the
player $A$ receives money from the player $B$, otherwise, the player
$A$ pays the player $B$. The process continues until the deck is
over. One is curious, as usual, about the winner and the amount of
his gain.

Let us now  look on such a random process more formally. Let $\pi =
[\pi_1,\pi_2,\pi_3,\ldots,\pi_{n+1}]$ denote a random permutation of
$[n+1]$. We rewrite it next in two-line notation as
\begin{equation}
\begin{pmatrix}
1&2&3& \dots& n+1\\\pi_1&\pi_2&\pi_3& \ldots &\pi_{n+1}
\end{pmatrix}
\end{equation}
and suppose that this table assigns a discrete "time" variable $l$
($l=1,2,.3,..., n+1$, upper line in the table) to each permutation
in the second line.
 We call, in a standard notation, as a "rise"
(or as a "descent") of the permutation $\pi$, such values of $l$ for
which $\pi_l < \pi_{l+1}$ ($\pi_l > \pi_{l+1}$).

Consider now a 1d lattice, place a walker at the origin at time
moment $l=0$, and let it move according to the
following rules:\\
-at time moment $l = 1$  the walker is moved one step to the right
if $\pi_1 < \pi_2$, i.e., $l = 1$ is a rise, or to the left if
$\pi_1
> \pi_2$, i.e $l = 1$ is a descent.\\
-at $l = 2$  the walker is moved to the right (left) if $\pi_2 <
\pi_3$ ($\pi_2 > \pi_3$, resp.) and etc.

Repeated $l$ times, this results in a random trajectory $X_l^{(n)}$,
 $(l = 1,2, \ldots , n)$, such that
\begin{equation}
\label{def} X_{l}^{(n)} = \sum_{k=1}^{l} s_k, \;\;\; s_k = {\rm
sgn}(\pi_{k+1}-\pi_{k}), \;\;\; {\rm sgn}(x) \equiv
\begin{cases}
+1& \text{if $x > 0$},\\
-1& \text{otherwise}.
\end{cases}
\end{equation}
Evidently, $X_{l}^{(n)}$ is just an amount of money the player $A$
won (or lost, if $X_{l}^{(n)} < 0$) up to the moment when the $l$-th
card is opened. Below we answer on a set of questions on various
statistical properties of random variable $X_{l}^{(n)}$.

\subsection{Probability Distribution of The End-Point of Trajectory $X_{l}^{(n)}$.}

Let ${\cal P}_n(X)$ denote the probability that the walker is at
site $X$ at time moment $l = n$.  Let ${\cal N}_{\uparrow}$ (${\cal
N}_{\downarrow}$) be the number of "rises" ("descents") in a given
permutation $\pi$. Evidently, the end-point $X_{n}$ of the walker's
trajectory is $X_{n} = {\cal N}_{\uparrow} - {\cal N}_{\downarrow}$.
Since ${\cal N}_{\uparrow} + {\cal N}_{\downarrow} = n$, we have
that $X_{n}  = 2 {\cal N}_{\uparrow} - n$ and hence, $X_{n}$ is
fixed by the number of rises in this permutation.

Number of permutations of $[n+1]$ with \textit{exactly} ${\cal
N}_{\uparrow}$ rises is given by the Eulerian number \cite{2}:
\begin{equation}
\label{en} \left \langle
\begin{matrix}
n+1 \\   {\cal N}_{\uparrow}
\end{matrix}
\right \rangle = \sum_{r=0}^{{\cal N}_{\uparrow} + 1} (-1)^r {n+2
\choose r} ({\cal N}_{\uparrow} + 1 - r)^{n+1}, \;\;\; {n+2 \choose
r} = \frac{(n+2)!}{r! (n + 2 - r)!}.
\end{equation}
Consequently,  ${\cal P}_{n}(X)$  is given explicitly by \cite{gr}
\begin{equation}
\label{eu}  {\cal P}_n(X) = \frac{[1 + (-1)^{X + n}]}{2 (n + 1)!}
\left \langle
\begin{matrix}
n+1 \\  \frac{X + n}{2}
\end{matrix}
\right \rangle.
\end{equation}
Several useful integral representations of the distribution function
${\cal P}_n(X)$ and of the corresponding lattice Green function were
also derived \cite{gr}. Expression in Eq.(\ref{euu}) may be cast
into the following form:
\begin{equation}
\label{int}  {\cal P}_n(X) = \frac{[1 + (-1)^{X + n}]}{\pi}
\int^{\infty}_0 \left(\frac{\sin(k)}{k}\right)^{n+2} \cos(X k) d k.
\end{equation}
Note that it looks \textit{almost} like usually encountered forms
for P\'olya walks \cite{hughes}, just the upper limit of integration
is infinity but not $"\pi"$; as a matter of fact, an integral
representation with the integration extending over the first
Brillouin zone \textit{only} obeys
\begin{eqnarray}
\label{br1}  {\cal P}_n(X) = \frac{(-1)^{n+1}}{(n+1)! \pi}
\int_{0}^{\pi}
 \left( \sin^{n+2}(k) \frac{d^{n+1}}{d k^{n+1}} \cot(k) \right) \cos(X k)
 dk,
\end{eqnarray}
and has a very different structure compared to that of P\'olya
walks. So does  the lattice Green function ${\cal G}(X,z)$,
associated with the result in Eq.(\ref{br1}):
\begin{eqnarray}
\label{br2}  {\cal G}(X,z) = \sum_{n=0}^{\infty} {\cal P}_n(X) z^n =
\frac{1}{\pi z} \int^{\pi}_0 \frac{\sin\left(z \sin(k)\right) \cos(X
k)}{\sin\left(k - z \sin(k)\right)} dk.
\end{eqnarray}
To establish the asymptotic form of   ${\cal P}_n(X)$ we go to the
limit $z \to 1^{-}$, and, inverting the expression  in
Eq.(\ref{br2}) in this limit, we find that ${\cal P}_{n}(X)$
converges to a \textit{normal} distribution:
\begin{equation}
\label{normal} {\cal P}_{n}(X) \sim  \left(\frac{3}{2 \pi
n}\right)^{1/2} \exp\left(- \frac{3 X^2}{2 n}\right), \;\;\;  n \to
\infty.
\end{equation}
Consequently, in this limit the correlations in the generator of the
walk - random permutations of $[n+1]$,  appear to be marginally
important; that is, they do not result in \textit{anomalous}
diffusion, but merely affect the "diffusion coefficient" making it
three times smaller than the diffusion coefficient of the standard
1D P\'olya walk.

\subsection{Correlations in
PGRW Trajectories.}

So, we have a symmetric random walk, which makes a move of unit
length at each moment of time with probability $1$,  but nonetheless
looses somehow two thirds of the diffusion coefficient. To elucidate
this puzzling question, we address now an "inverse" problem: given
the distribution ${\cal P}_n(X)$, we aim to determine two- and
four-point correlations in the rise-and-descent sequences and
consequently, in the PGRW trajectories.

For the Fourier-transformed ${\cal P}_n(X)$ we find
\begin{eqnarray}
\label{tt} \tilde{{\cal P}}_n(k) = \sum_{X = - n}^{n} \exp\left(i k
X\right) {\cal P}_n(X) = \frac{(2 i)^{n+2} \sin^{n+2}(k)}{(n + 1)!}
{\rm Li}_{-n-1}\Big(\exp(- 2 i k)\Big),
\end{eqnarray}
${\rm Li}_{-n-1}(y)$ being a poly-logarithm function. Eq.(\ref{tt})
yields
\begin{equation}
\label{x2} \Big \langle X_n^2 \Big \rangle \equiv \frac{1}{3} \Big(n
+ 2 - 2 \delta_{n,0}\Big) \;\;\; \text{and} \;\;\; \Big \langle
X_n^4 \Big \rangle \equiv \frac{1}{15} \Big((5 n + 8) ( n + 2) -16
\delta_{n,0} - 24 \delta_{n,1} + 8 \delta_{n,2} \Big),
\end{equation}
where $\delta_{i,j}$ is the  Kronecker delta, such that their
generating functions are
\begin{eqnarray}
{\cal X}^{(2)}(z) &=& \sum_{n=0}^{\infty} \left\langle X_{n}^2
\right\rangle z^{n} = \frac{z \left(3 - 2 z\right)}{3 \left(1 -
z\right)^2}, \nonumber\\ {\cal X}^{(4)}(z) &=& \sum_{n=0}^{\infty}
\left\langle X_{n}^4 \right\rangle z^{n} = \frac{z \left(35 z  + 15
- 80 z^2 + 48 z^3 - 8 z^4\right)}{15 (1 - z)^3}.
\end{eqnarray}
On the other hand, the second and the fourth moments of the walker's
displacement obey
\begin{eqnarray}
\label{q} \left\langle X_{n}^2 \right\rangle &=&  \left\langle
\left({\cal N}_{\uparrow} - {\cal N}_{\downarrow} \right)^2
\right\rangle  =  \left \langle \left[\sum_{l=1}^{n} s_l \right]^{2}
\right\rangle =
 n + 2 \sum_{j_{1} = 1}^{n - 1} \sum_{j_2 = j_1 + 1}^{n} {\cal
 C}_{j_1,j_2}^{(2)}, \nonumber\\
\left\langle X_{n}^4 \right\rangle &=&  n +  \frac{n (n - 1)}{2}
\theta(n-2) + \Big(8 (n - 1)  \theta(n - 2) +   12 \left(n
(n - 3) + 2 \right) \theta(n-3)\Big) {\cal C}^{(2)}(1) + \nonumber\\
&+& 24 \sum_{m = 1}^{n - 3} \Big(n - 2 - m\Big) {\cal C}^{(4)}(m)
\theta(n-4),  \;\;\; \theta(x) \equiv
\begin{cases}
+1& \text{if $x > 0$},\\
0& \text{otherwise}.
\end{cases}
\end{eqnarray}
where ${\cal C}^{(2)}(m) = {\cal C}_{j_1,j_{m+1}}^{(2)}= \langle
s_{j_1} s_{j_{m+1}} \rangle$ is  the two-point correlation function
of the rise-and-descent sequence, while ${\cal C}^{(4)}(m) = \langle
s_{j_1} s_{j_1+1} s_{j_1+m+1} s_{j_1+m+2} \rangle$ defines the
four-point correlation function. Note that $k$-point correlations
with $k$ - odd vanish \cite{gr} and thus ${\cal C}^{(4)}(m)$ is an
only non-vanishing form of four-point correlations.

Multiplying both sides of Eqs.(\ref{q}) by $z^n$ and performing
summation, we find that
\begin{eqnarray}
\label{as} {\cal C}^{(2)}(z) &=& \sum_{m=1}^{\infty} z^{m} {\cal
C}^{(2)}(m) = \frac{(1-z)^2}{2 z} {\cal X}^{(2)}(z)  -\frac{1}{2} =
-
\frac{z}{3}, \nonumber\\
{\cal C}^{(4)}(z) &=& \sum_{m=1}^{\infty} z^{m} {\cal C}^{(4)}(m) =
\frac{(1-z)^2}{24 z^3} {\cal X}^{(4)}(z)  + \frac{16 z^2 - 7 z -
3}{72 z^2 (1-z)} = \nonumber\\
&=& \frac{2}{15} z + \frac{1}{9} \left(z^2 + z^3 + z^4 +
\cdots\right),
\end{eqnarray}
which imply
\begin{equation}
\label{zzz} {\cal C}^{(2)}(m) =
\begin{cases}
 -1/3& \text{$m = 1$}\\
 0& \text{$m \geq 2$}
\end{cases}
\;\;\; {\cal C}^{(4)}(m) =
\begin{cases}
 2/15& \text{if $m = 1$}\\
 1/9& \text{if $m \geq 2$}
\end{cases}
\end{equation}
Equations (\ref{zzz}) signify that for $m \geq 2$ the two-point
correlations ${\cal C}^{(2)}(m = |j_2 - j_1|)$ in the
\text{rise-and-descent} sequences decouple into the product
$<s_{j_1}> <s_{j_2}>$ and hence, vanish. Consequently, two-point
correlations extend to the nearest neighbors only. The four-point
correlations decouple  ${\cal C}^{(4)}(m) = {\cal C}^{(2)}(1) {\cal
C}^{(2)}(1)$ for $m \geq 2$.

Equations (\ref{zzz}) enable us to calculate explicitly the
probabilities of different configurations of two, three and four
rises and descents, which are listed below
\begin{equation}
\label{o1}  p_{\uparrow, \uparrow}(m) = p_{\downarrow,
\downarrow}(m)
 =
\begin{cases}
 1/6 & \text{$m = 1$}\\
 1/4 & \text{$m \geq 2$}
\end{cases}
\;\;\; p_{\uparrow, \downarrow}(m) = p_{\downarrow, \uparrow}(m)
 =
\begin{cases}
 1/3& \text{$m = 1$}\\
 1/4& \text{$m \geq 2$}
\end{cases}
\nonumber\\
\end{equation}
Hence, configurations with two neighboring rises or two neighboring
descents have a lower probability than mixed rise-descent or
descent-rise sequences; in other words, rises "repel" each other and
"attract" descents. Further on, we have:
\begin{equation}
\label{p0} p_{\uparrow \uparrow, \downarrow}(m) =
\begin{cases}
 1/4 & \text{$m = 1$}\\
 1/12 & \text{$m \geq 2$}
\end{cases}
\;\;\; p_{\uparrow \uparrow, \uparrow}(m)  =
\begin{cases}
 1/24& \text{$m = 1$}\\
 \frac{1}{12}& \text{$m \geq 2$}
\end{cases}
\;\;\; p_{\uparrow \uparrow, \uparrow \uparrow}(m) =
\begin{cases}
 1/120 & \text{$m = 1$}\\
 1/36  & \text{$m \geq 2$}
\end{cases}
\nonumber\\
\end{equation}
\begin{equation}
\label{p2} p_{\uparrow \uparrow,  \downarrow \uparrow}(m) =
p_{\uparrow \downarrow,  \uparrow \uparrow}(m) =
\begin{cases}
 3/40 & \text{$m = 1$}\\
 1/18 & \text{$m \geq 2$}
\end{cases}
\;\;\; p_{\uparrow \uparrow,  \uparrow \downarrow}(m) =
p_{\downarrow \uparrow,  \uparrow \uparrow}(m) =
\begin{cases}
 1/30  & \text{$m = 1$}\\
 1/18  & \text{$m \geq 2$}
\end{cases}
\nonumber\\
\end{equation}
\begin{equation}
\label{p3} p_{\uparrow \downarrow,  \uparrow \downarrow}(m) =
\begin{cases}
 2/15   & \text{$m = 1$}\\
 1/9    & \text{$m \geq 2$}
\end{cases}
\;\;\; p_{\uparrow \uparrow, \downarrow \downarrow}(m) =
\begin{cases}
 1/20    & \text{$m = 1$}\\
 1/36    & \text{$m \geq 2$}
\end{cases}
\;\;\; p_{\downarrow \uparrow, \uparrow \downarrow}(m) =
\begin{cases}
 11/120  & \text{$m = 1$}\\
 1/9     & \text{$m \geq 2$}
\end{cases}
\nonumber\\
\end{equation}
Analyzing these results, we notice that the probabilities of
different rise-and-descent sequences depend not only on the number
of rises or descents, but also on their order within the sequence.
\begin{figure}[ht]
\begin{center}
\includegraphics*[scale=0.5, angle=0]{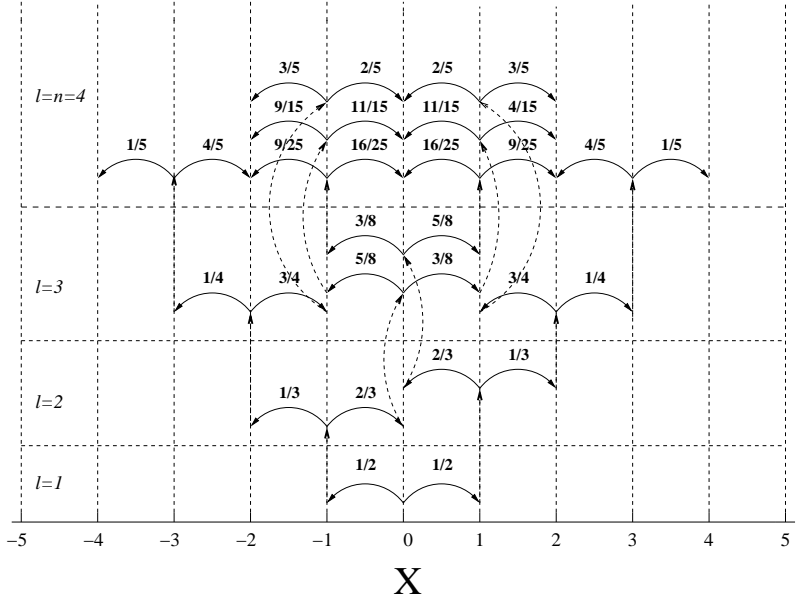}
\caption{\label{Fig1} {A set of all possible PGRW trajectories
$X_l^{(n)}$ for $n = 4$. Numbers above the solid arcs with arrows
indicate the corresponding transition probabilities. Dashed-lines
with arrows connect the trajectories for different values of $l$.}}
\end{center}
\end{figure}
Correspondingly, reconstructing the PGRW trajectories $X^{(n)}_l$
with $n = 4$ (see Fig.1) we notice that the PGRW transition
probabilities depend not only on the number of steps to the right or
to the left, which the walker has already made, but also on their
order. In other words, the PGRW represents a genuine non-Markovian
process with a memory.

Finally, for $k$-point correlation functions ${\cal
C}^{(k)}_{j_1,..,j_k} = \langle s_{j_1} s_{j_2} s_{j_3} \cdots
s_{j_k} \rangle$ of the rise-and-descent sequences we find the
following behavior \cite{gr}:  ${\cal C}_{j_1,..,j_k}^{(k)}$
factorizes automatically into a product of the corresponding
correlation functions of the consecutive subsequences, in which all
$j_k$ differ by unity, as soon as any of the distances $j_{k+1} -
j_k $ exceeds unity. On the other hand, the correlation function
${\cal C}^{(k)} = {\cal C}^{(k)}_{j,j+1,..,j+k}$ of a consecutive
sequence of arbitrary order $k$  can be obtained exactly. One has
that ${\cal C}^{(k)}$ obey the following recursion \cite{gr}:
\begin{eqnarray}
{\cal C}^{(k)} = \sum_{p=0}^{k-1} \frac{(-1)^k 2^k}{(k+1)!} {\cal
C}^{(k-1-p)}, \;\;\; {\cal C}^{(0)} = 1,
\end{eqnarray}
which implies that ${\cal C}^{(k)}$ are related to the tangent
numbers \cite{foata}, and are given explicitly by
\begin{equation}
\label{dddf} {\cal C}^{(k)} = \frac{(-1)^k 2^{k+2} \left(2^{k+2} -
1\right)}{(k + 2)!} B_{k+2},
\end{equation}
where $B_k$ are the Bernoulli numbers. Note that since the Bernoulli
numbers equal zero for $k$ odd,  correlation functions of odd order
vanish, i.e., ${\cal C}^{(2 k + 1)} \equiv 0$. Eq.(\ref{dddf}) also
implies that  for $k \gg 1$, the $k$-point correlation functions
${\cal C}^{(k)}$ ($k$ is even) decay as ${\cal C}^{(k)} \sim 2
(-1)^{k/2} \left(2/\pi\right)^{k+2}$.

\subsection{Probability
Distribution of ${\bf X^{(n)}_l}$ for ${\bf l < n}$.}

To determine the structure of excursions $X^{(n)}_l$ of the PGRW we
adapt a method proposed by Hammersley \cite{hammersley} in his
analysis of the expected length of the longest increasing
subsequence. The basic idea behind this approach is to build
recursively an auxiliary \textit{Markovian} stochastic process
$Y_l$, which is distributed exactly as $X^{(n)}_l$.

At each time step $l$, we define a real valued random variable
$x_{l+l}$, uniformly distributed in $[0,1]$. Further on, we consider
a random walk on a one-dimensional lattice of integers whose
trajectory $Y_l$ is constructed according to the following
step-by-step process: at each time moment $l$ a point-like particle
is created at position $x_{l+1}$. If $x_{l+1} > x_{l}$, a walker is
moved one step to the right; otherwise, it is moved one step to the
left. The trajectory $Y_l$ is then given by $ Y_l=\sum_{k=1}^{l}
{\rm sgn}(x_{k+1}-x_k)$, where ${\rm sgn}(x)$ is defined in
Eq.(\ref{def}).

We note that the joint process $(x_{l+1},Y_l)$, and therefore $Y_l$,
are \textit{Markovian} since
 they depend only on $(x_{l},Y_{l-1})$. Note also that $Y_l$ is the sum of
{\it correlated} random variables; hence, one has to be cautious
when applying central limit theorems. A central limit theorem indeed
holds for the Markovian process  $Y_l$, but the summation rule for
the variance is not valid.

Two following results have been proven in Ref.\cite{gr} concerning
the relation between this recursively constructed Markovian process
$Y_l$ and the PGRW:\\ (a) the probability $P(Y_l = X)$ that the
trajectory $Y_l$ of the auxiliary process appears at site $X$ at
time moment $l$ is equal to the probability ${\cal P}_l(X) =
P(X_l^{(l)}= X)$ that the end-point $X_l$ of the PGRW trajectory
$X_l^{(l)}$ generated by a given permutation of $[l+1]$ appears at
site $X$.\\
(b) the probability $P(X_l^{(n)} = X)$ that the PGRW trajectory
$X_l^{(n)}$ at intermediate time moment $l = 1,2,3, \ldots,n$
appears at site $X$ is equal to the probability $P(Y_l = X)$ that
the trajectory of the auxiliary process $Y_l$ appears at time moment
$l$ at site $X$. In other words, one has
\begin{equation}
\label{euu}  P(Y_l = X)= P(X_l^{(n)} = X) = {\cal P}_l(X) = \frac{[1
+ (-1)^{X + l}]}{2 (l + 1)!} \left \langle
\begin{matrix}
l+1 \\  \frac{X + l}{2}.
\end{matrix}
\right \rangle.
\end{equation}
Note that distribution of any intermediate point $X_l^{(n)}$ of the
PGRW trajectory generated by permutations of a sequence of length
$n+1$ depends on $l$ but is independent of $n$.

\subsection{Probability Measure of PGRW Trajectories.}
\label{sss}

The equivalence of the processes $Y_l$ and $X_l^{(n)}$
 allows to determine
the probability measure of any given trajectory. We note that in the
permutation language, this problem amounts to the calculation of the
number of permutations with a given rise-and-descent sequence and
has been already discussed using an elaborated combinatorial
approach in Refs.\cite{macmahon,carlitz,niven}. A novel solution of
this problem has been proposed in Ref.\cite{gr}, which expressed the
probability measure of any given PGRW trajectory (or of some part of
it) as a chain of iterated integrals.

Consider a given rise-and-descent sequence $\alpha(k)$ of length $k$
of the form:
$$
\alpha(k)=\left(\begin{array}{ccccc} 1 & 2 & 3 & ... & k \\
a_1 & a_2 & a_3 & ... & a_{k}
\end{array} \right).
$$
where $a_l$ can take either of two symbolic values --- $\uparrow$ or
$\downarrow$. Consequently, the first line in the table is the
running index $l$ which indicates position along the permutation,
while the second line shows what we have at this position - a rise
or a descent. Assign next to each symbol at position $l$ an integral
operator; $I_l(\uparrow)$ for a rise ($\uparrow$) and
$I_l(\downarrow)$ for a descent ($\downarrow$):
\begin{equation} \hat{I}_l(\uparrow)= \int_{x_{l-1}}^1 dx_{l} \qquad \mbox{and}
\qquad \hat{I}_l(\downarrow)=\int_0^{x_{l-1}} dx_{l}.
\label{eq:cor1} \end{equation} Define next a characteristic
polynomial $Q(x, \alpha(k))$ as an "ordered'' product \cite{gr}:
\begin{equation} Q(x,\alpha(k))=\; : \hspace{-4pt} \prod_{l=1}^k
\hspace{-4pt} : \; \hat{I}_l(a_l) \cdot 1 \label{eq:cor2},
\end{equation} where $a_l=\{\uparrow,\downarrow\}$ for $l=1,...,k$.
The  probability measure  $P(\alpha(k))$ of this given
rise-and-descent sequence $\alpha(k)$ in the ensemble of all equally
likely permutations is then given by \begin{equation}
P(\alpha(k))=\int_0^1 Q(x,\alpha(k)) dx. \label{eq:cor2a}
\end{equation}

In may be expedient to illustrate this formal consideration on a
particular example. Let a given rise-and-descent sequence be of the
form $\{\uparrow,\uparrow,\downarrow,\uparrow,\uparrow\}$. For this
sequence, the characteristic polynomial $Q(x,\alpha(5))$ and the
probability of this particular configuration obey
\begin{eqnarray}
&&Q(x,\alpha(5))= \hat{I}_1(\uparrow)\,
\hat{I}_2(\uparrow)\,\hat{I}_3(\downarrow)\, \hat{I}_4(\uparrow)\,
\hat{I}_5(\uparrow) \cdot 1 = \int\limits_x^1 dx_1
\int\limits_{x_{1}}^{1} dx_2 \int\limits_{0}^{x_{2}} dx_3
\int\limits_{x_3}^1 dx_4 \int\limits_{x_{4}}^{1} dx_5 \cdot 1 =
\nonumber\\ &=& \frac{3}{40} -\frac{x}{8}+\frac{x^{3}}{12}-
\frac{x^{4}}{24}+\frac{x^{5}}{120} \;\;\; \text{and} \;\;\;
P(\alpha(5))=\int_0^1 Q(x,\alpha(5))\,dx =\frac{19}{720}.
\end{eqnarray}

Another way to look on the problem of calculation of the measure of
a given PGRF trajectory is to use the results of Niven determining
the number of permutations with a given rise-and-descent sequence
\cite{niven}. Following Niven, consider a fixed \text{up-and-down}
arrow sequence $\alpha(k)$ of length $k$ and denote by $l_1, l_2,
\ldots, l_r$ the positions of downarrows (descents), $r$ being the
total number of downarrows along the sequence.  A question now is to
calculate the \textit{number} $N(X_l^{(n)})$ of permutations
generating a given \text{up-and-down} sequence  (or, in our
language, a given trajectory $X_l^{(n)}$). Combinatorial arguments
show that $N(X_l^{(n)})$ equals the determinant of a matrix of order
$r+1$ whose elements $a_{i,j}$ (where $i$ stands for the row, while
$j$ - for the column) are binomial coefficients $l_i \choose
l_{j-1}$, where $l_0=0$ and $\ l_{r+1}=k+1$ \cite{niven}.
Consequently, an alternative expression for the probability
$P(\alpha(k))$ may be written down as
\begin{equation}
\label{det} P(\alpha(k))= \frac{1}{(k + 1)!} {\rm det}
\begin{pmatrix}
\displaystyle
1 & 1& \displaystyle l_1 \choose l_2 & \displaystyle l_1 \choose l_3 & \ldots &  \displaystyle l_1 \choose l_r \\
1 & \displaystyle l_2 \choose l_1 & 1 & \displaystyle l_2 \choose l_3 & \ldots &  \displaystyle l_2 \choose l_r\\
1 & \displaystyle l_3 \choose l_1 & \displaystyle l_3 \choose l_2 & 1 & \dots & \displaystyle l_3 \choose l_r\\
&&&\ldots &\\
1 &  \displaystyle l_{r+1} \choose l_1 & \displaystyle l_{r+1}
\choose l_2 & \displaystyle l_{r+1} \choose l_3 & \dots &
\displaystyle l_{r+1} \choose _r
\end{pmatrix}.
\end{equation}
One may readily verify that both Eq.(\ref{eq:cor2a}) and
Eq.(\ref{det}) reproduce our
 earlier results
 determining the probabilities of different four step
trajectories. Note also that the probability measure defined by
Eq.(\ref{eq:cor2a}) or Eq.(\ref{det}) is not homogeneous, contrary
to the measure of the standard P\'olya walk.

\subsection{Distribution of The Number of "U-Turns" of PGRW
Trajectories.}

In this subsection we study an important measure of how scrambled
the PGRW trajectories are. This measure is the number ${\cal N}$ of
the "U-turns" of an $n$-step PGRW trajectory, i.e. the number of
times when the walker changes the direction of its motion.

In the permutation language, each turn to the left (right), when the
walker making a jump to the right (left) at time moment $l$ jumps to
the left (right) at the next time moment $l+1$ corresponds  to a
peak $\uparrow \downarrow$ (a through $\downarrow \uparrow$) of a
given permutation $\pi$, i.e. a sequence $\pi_{l} < \pi_{l+1} >
\pi_{l+2}$ ($\pi_{l} > \pi_{l+1} < \pi_{l+2}$). Consequently, the
distribution function ${\cal P}({\cal N},n)$ of the number of the
"U-turns" of the PGRW trajectory, (i.e. the probability that an
$n$-step PGRW trajectory has exactly ${\cal N}$ turns), is just the
distribution of the sum of peaks and through.

Let ${\cal N}_p$ and ${\cal N}_t$ denote the number of peaks  and
through  in a given random permutation $[n+1]$, respectively. These
realization-dependent numbers can be written down as
\begin{equation} {\cal N}_p = \frac{1}{4} \sum_{l = 1}^{n-1} \left(1 +
s_l\right) \left(1 - s_{l+1}\right), \;\;\; {\cal N}_t = \frac{1}{4}
\sum_{l = 1}^{n-1} \left(1 - s_l\right) \left(1 + s_{l+1}\right),
\end{equation}
such that total number ${\cal N}$ of U-turns  of the PGRW trajectory
is given by
\begin{equation} {\cal N} = {\cal N}_p + {\cal N}_t = \frac{1}{2}
\sum_{l = 1}^{n-1} \left(1 - s_l s_{l+1}\right)
\end{equation}
One finds then that the characteristic function ${\cal Z}_n(k)$ of
${\cal N}$ is a polynomial in $\tanh(i k/2)$ \cite{gr}:
\begin{eqnarray}
\label{partition}  &&{\cal Z}_n(k) = \Big\langle \exp\left(i k {\cal
N}\right) \Big \rangle = \exp\left( \frac{i k (n - 1)}{2} \right)
\Big\langle \exp\left(- \frac{i k}{2} \sum_{l=1}^{n-1} s_l
s_{l+1}\right) \Big \rangle =  \left(\frac{1+e^{i k}}{2}\right)^{n-1} \times   \nonumber\\
&\times&  \sum_{l=0}^{[n/2]} (-1)^l \left( \sum_{1 m_1 + 2 m_2 +
\ldots + l m_l = l} { n - 2l + 1 \choose \sum_{j=1}^l m_j }
\frac{\left(\sum_{j=1}^l m_j\right)!}{m_1! m_2! \ldots m_l!}
\prod_{j = 1}^{l} \left( {\cal C}^{(2 j)} \right)^{m_j}\right) \,
\tanh^l(\frac{i k}{2}), \nonumber
\end{eqnarray}
where ${\cal C}^{(2 j)}$ obey Eq.(2.16). The generating function of
${\cal Z}_n(k)$ is given by
\begin{eqnarray}
{\cal Z}(k,z) &=& \sum_{n = 2}^{\infty} {\cal Z}_n(k) \; z^n = -
 \frac{4}{\left(1 + e^{i k}\right)^2 z} - \frac{2}{1 + e^{i k}} - z + \nonumber\\
&+&\frac{4}{\left(1 + e^{i k}\right)^2 z} \left[1 - \frac{\left(1 +
e^{i k}\right) z}{2} \sum_{j = 0}^{\infty}{\cal C}^{(2 j)} \left(
\frac{\left(1 - e^{2 i k}\right) z^2}{4}\right)^{j}\right]^{-1} =
\nonumber\\
&=&  \frac{4}{\left(1 + e^{i k}\right)^2 z} \left[\left(\frac{1-e^{i
k}}{1 + e^{i k}}\right)^{1/2} \coth\left(\left(1 - e^{2 i
k}\right)^{1/2} \frac{z}{2}\right)-1\right]^{-1} - \frac{2}{1 + e^{i
k}} - z \nonumber
\end{eqnarray}
Turning next to the limit $z \to 1^{-}$ and inverting ${\cal
Z}(k,z)$ with respect to $k$ and $z$, we find that in the asymptotic
limit $n \to \infty$, the distribution function ${\cal P}({\cal
N},n)$ of the number of the "U-turns" in the PGRW trajectory
converges to a normal distribution: \begin{equation} {\cal P}({\cal
N},n) \sim \frac{3}{4} \Big(\frac{5}{\pi n}\Big)^{1/2} \exp\left(-
\frac{45 \Big({\cal N} -  \frac{2}{3} n\Big)^2}{16 n}\right),
\end{equation} with mean
value $2 n/3$ and variance $\sigma^2 = 8 n/45$.

\subsection{Diffusion Limit}

Consider finally a continuous space and time version of the PGRW in
the diffusion limit. In Ref.\cite{gr}, the following approach has
been developed:\\
 Define first the polynomial: $V^{(l)}(x,Y)=\sum_{Y_l = Y} Q(x,\alpha(l))$,
 where the polynomial
$Q(x,\alpha(l))$ has been determined in Eq.(\ref{eq:cor2}) and the
sum extends over all $l$-steps trajectories $Y_l$ starting at zero
and ending at the fixed point $Y$. Note next that one has ${\cal
P}_l(Y)=\left<V^{(l)}(x,Y)\right>_{\{x_l\}}$. Now, for the
polynomials $V^{(l)}(x,Y)$ one obtains, by counting all possible
trajectories $Y_l$ starting from zero and ending at the fixed point
$Y$, the following "evolution" equation: \begin{equation}
\label{master} V^{(l+1)}(x,Y)=\hat{I}_l(\uparrow) \cdot
V^{(l)}(x,Y-1)+\hat{I}_l(\downarrow) \cdot V^{(l)}(x,Y+1).
\end{equation}
Taking next advantage of the established equivalence between the
processes $Y_l$ and $X_l^{(n)}$, we can rewrite the last equation,
upon averaging it over the distribution of variables $\{x_l\}$, as
\begin{equation} \label{master2} {\cal P}_{l+1}(Y)=\frac{(l-Y+4)}{2(l+1)}{\cal
P}_{l}(Y-1)+\frac{(l+Y)}{2(l+1)}{\cal P}_{l}(Y+1),
\end{equation}
which represents the desired evolution equation for ${\cal
P}_{l}(Y)$ in discrete space and time.

We turn next to the diffusion limit. Introducing $y = a Y$ and  $t =
\tau l$ variables, where $a $ and $\tau$ define characteristic space
and time scales, we turn to the limit $a, \tau \to 0$, supposing
that the ratio $a^2/\tau$ remains fixed and determines the diffusion
coefficient $D_0 = a^2/ 2 \tau$. In this limit, Eq.(\ref{master2})
becomes \begin{equation}\label{fokker} \frac{\partial}{\partial
t}{\cal P}(y,t)=\frac{\partial}{\partial y}\left(\frac{y}{t}{\cal
P}(y,t)\right)+D_0\frac{\partial^2}{\partial y^2}{\cal P}(y,t)
\end{equation}
 Note that the resulting continuous space and time equation is
of the Fokker-Planck type; it has a constant diffusion coefficient
and a negative drift term which, similarly to the Ornstein-Uhlenbeck
process, grows linearly with $y$, but the amplitude of the drift
decays in proportion to the first inverse power of time, which
signifies that the process $Y_l$ eventually delocalizes. The Green's
function solution of Eq.(\ref{fokker}),  remarkably, is a normal
distribution \begin{equation} {\cal P}(y,t)=\sqrt{\frac{3}{4\pi D_0
t}} \exp\left(- \frac{3 y^2}{4 D_0t}\right), \end{equation}
 which is consistent with the large-$n$
limit of the discrete process derived in Eq.(\ref{normal}).

\section{Random Surfaces Generated by Random Permutations.}

\begin{figure}[ht]
\begin{center}
\includegraphics*[scale=0.6, angle=0]{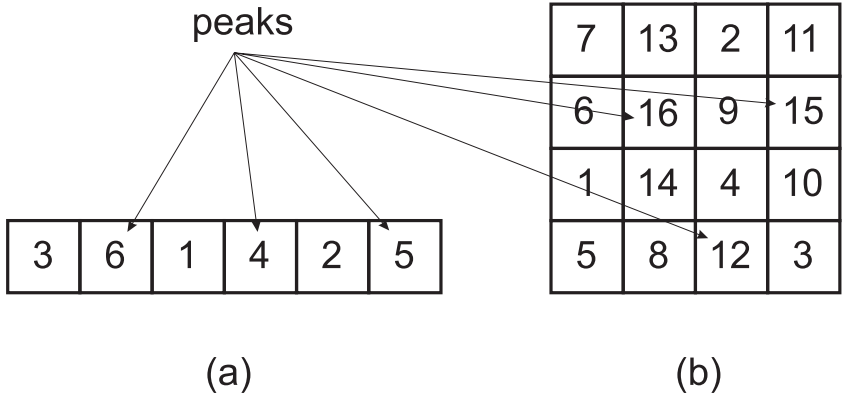}
\caption{\label{Fig2} {(a) One--dimensional ($L=6$) and (b)
two-dimensional square ($L=4 \times 4$) lattices with periodic
boundary conditions on sites of which we distribute numbers drawn
from the set $1,2,3, \ldots, L$. These numbers determine the local
heights of the surface. The sites with numbers bigger than those
appearing on the neighboring sites are refereed to as the "peaks".}}
\end{center}
\end{figure}

Consider a one--dimensional lattice containing $L$ sites on which we
distribute at random numbers drawn from the set $1,2,3, \ldots,L$,
(see Fig.\ref{Fig2}). Suppose that a number appearing at the site
$j$ determines the local height of the surface. Further on, we
denote a local "peak" as a site $j$ the number at which exceeds the
numbers appearing at two neighboring sites. Generalization to
two--dimensional square lattice with $L = m \times m$, ($m$ is an
integer) sites is straightforward (see Fig.\ref{Fig2}): the only
difference here is that we call as local surface peaks such sites
$j$ the numbers at which are greater than numbers appearing at four
adjacent sites.

For these models, our goal is to evaluate the probability $P(M,L)$
that the surface created in such a way has exactly $M$ peaks on a
lattice containing $L$ sites \cite{hiv}. In one dimension this can
be done exactly and provides also a distribution function of the
number of right U-turns of the PGRW trajectories. Here we are also
able to calculate the "correlation function" $p(l)$ defining the
conditional probability that two peaks are separated by the interval
$l$ under the condition that the interval $l$ does not contain other
peaks. In 2D, determining exactly first three cumulants of $P(M,L)$
we define its asymptotic form  using expansion in the Edgeworth
series \cite{edg} and show that it converges to a normal
distribution as $L \to \infty$ \cite{hiv}. For 2D model, we will
also discuss some surprising cooperative behavior of peaks.

\subsection{Probability Distribution $P(M,L)$ in 1D Case
}

In one dimension, the probability $P(M,L)$ of having $M$ surface
peaks on a lattice containing $L$ sites can be calculated exactly
through its generating function \cite{hiv}:
\begin{equation} \label{mmm} W(s,L) = L! \sum_{M=0}^{\infty} s^{M+1}
P(M,L),
\end{equation}
where
\begin{equation}
 W(s,L) =
\left(\frac{s}{1-\sqrt{1-s}}\right)^{L+1} \sum_{M=0}^{\infty}
\left(\frac{(1 - \sqrt{1-s})^2}{s}\right)^{M+1} \left \langle
\begin{matrix}
L \\  M - 1
\end{matrix}
\right \rangle .
\end{equation}
Inverting (\ref{mmm}), we find the following exact expression for
$P(M,L)$:
\begin{equation}
 P(M,L) = \frac{2^{L+2}}{L!} \sum_{l=1}^M
(-1)^{M-l} \left(\begin{array}{c} \frac{L+1}{2} \\ M - l
\end{array} \right)
 \sum_{m=1}^l
\frac{m^{L+1}}{(l-m)! (m+l)!} \label{eq:prob_exact}
\end{equation}
One finds then that in the limit $L \to \infty$, the probability
$P(M,L)$ converges to a normal distribution: \begin{equation}
P(M,L)\sim \frac{3}{2}\sqrt{\frac{5}{\pi L}}
\exp\left\{-\frac{45(M-\frac{1}{3}L)^2}{4L}\right\} \label{eq:3b}
\end{equation}
with mean $L/3$ and variance $\sigma^2 = 2 L/45$. Note that this
result is consistent with the distribution of the number of U-turns
of the PGRW trajectories, Eq.(2.25).

\subsection{Conditional Probability $p(l)$ of Two Peaks Separated by Distance $l$}

We aim now at evaluating the conditional probability $p(l)$ of
having two peaks separated by a distance $l$, under the condition
that there are no peaks (i.e. sequences $\uparrow\, \downarrow$) on
the interval between these peaks. Using the approach put forth in
Section 2.4, we have that this probability obeys
\begin{equation} p(l) = \int^1_0 dx \sum Q(x,\alpha) \label{s}, \end{equation}
 where the sum
is taken over all possible peak-avoiding rise-and-descent patterns
of length $l$ between of two peaks, while $Q(x,\alpha)$ denote the
$Q$--polynomials corresponding to each given configuration.

There are several possible peak-avoiding rise-and-descent sequences
contributing to such a probability. These sequences are depicted in
Fig.3.

\begin{figure}[ht]
\begin{center}
\includegraphics*[scale=0.5, angle=0]{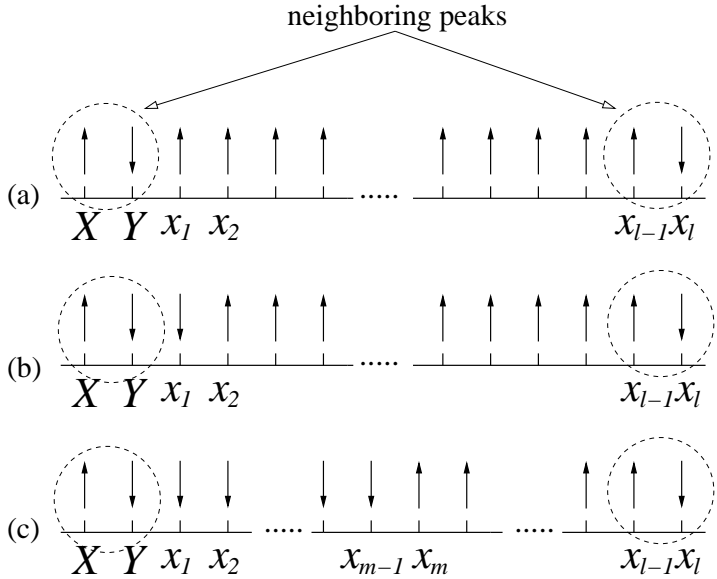}
\caption{\label{Fig3} {Rise-and-descent patterns contributing to the
conditional probability of having two closest peaks at distance $l$
apart from each other. Configuration (a) has only one descent
between two peaks. Configuration (b) has two descents following the
first peak, i.e. a "through" at $x_1$, and (c) presents a
generalization of (b) over configurations having $m$ descents, ($m =
1,2, \ldots, l$), after the first peak which are followed by $l-m$
rises, i.e. a "through" at $x_{m-1}$.}}
\end{center}
\end{figure}

The $Q$-polynomial associated with  the general configuration (c) in
Fig.3  is given by \begin{equation} \label{eq:qc} Q^c(x)  =
\int^{1}_x dX \int^{X}_0 dY \int^Y_{0} dx_1 \int^{ x_1}_0 dx_2
\ldots \int^{ x_{m-2}}_0 dx_{m-1} \int^1_{ x_{m-1}} dx_{m} \ldots
\int^1_{ x_{l-2}} dx_{l-1} \int^{ x_{l-1}}_0 dx_{l}
\end{equation} Calculating this integral recursively and summing
over $m$, $m=1,2,\ldots,l$, we find that the generating function of
the probability $p(l)$ obeys
\begin{eqnarray}
F(z) = \sum_{l=2} p(l) z^l &=& \frac{(z-1)^{2}}{2z^{3}}
e^{2z}+\left( \frac{1}{3}+\frac{1}{2z}-\frac{1}{2z^{3}} \right) =
\nonumber\\ &=& \frac{2}{15}z^{2}+ \frac{1}{9}z^{3}+
\frac{2}{35}z^{4}+ \frac{1}{45}z^{5}+ \frac{4}{567}z^{6}+\cdots
\end{eqnarray}
Inverting the latter expression, we find that $p(l)$ is determined
explicitly by
\begin{equation} p(l)=\frac{1}{2}\left( \frac{2^{l+1}}{(l+1)!}
-2\frac{2^{l+2}}{(l+2)!} +\frac{2^{l+3}}{(l+3)!}
\right)=2^{l}\frac{(l-1)(l+2)}{(l+3)!}, \label{eq:p(l)}
\end{equation} Note also that $p(l)$ in eq.(\ref{eq:p(l)}) coincides
with the distribution function of the distance between two "weak"
bonds obtained by Derrida and Gardner \cite{derrida2} in their
analysis of the number of metastable states in a one-dimensional
Ising spin glass at zero temperature.

\subsection{Probability Distribution $P(M,L)$ in 2D Case}

We turn now to calculation of the probability $P(M,L)$ of having $M$
peaks on a two-dimensional square lattice containing $m \times m =
L$ sites. We linearly order lattice sites by index $j$, $j=1,2,
\ldots, L$, in the same way as an electron beam highlights the TV
screen. Note that in such a representation a site $j$ is a peak if
and only if $\pi_j$ is simultaneously larger than $\pi_{j-1},
\pi_{j+1}, \pi_{j-m}$ and $ \pi_{j+m}$.

\begin{figure}[!htb]
\label{Fig4}
\begin{center}
\centerline{\hspace*{.2\textwidth}(a)\hspace*{.37\textwidth}(b)
\hfill}
\begin{minipage}[c]{.49\textwidth}
\includegraphics[width = 0.9\textwidth,height=0.9\textwidth]{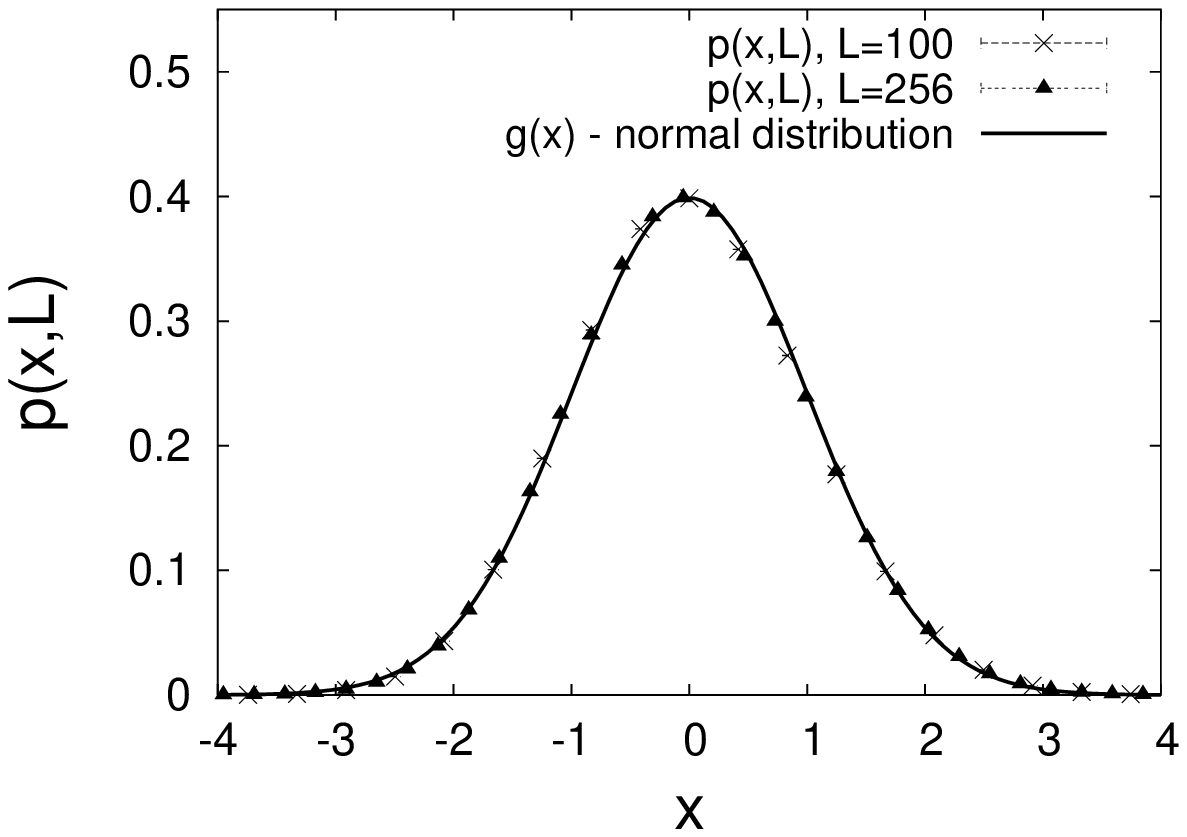}
\end{minipage}
\begin{minipage}[c]{.49\textwidth}
\includegraphics[width = 0.9\textwidth,height=0.9\textwidth]{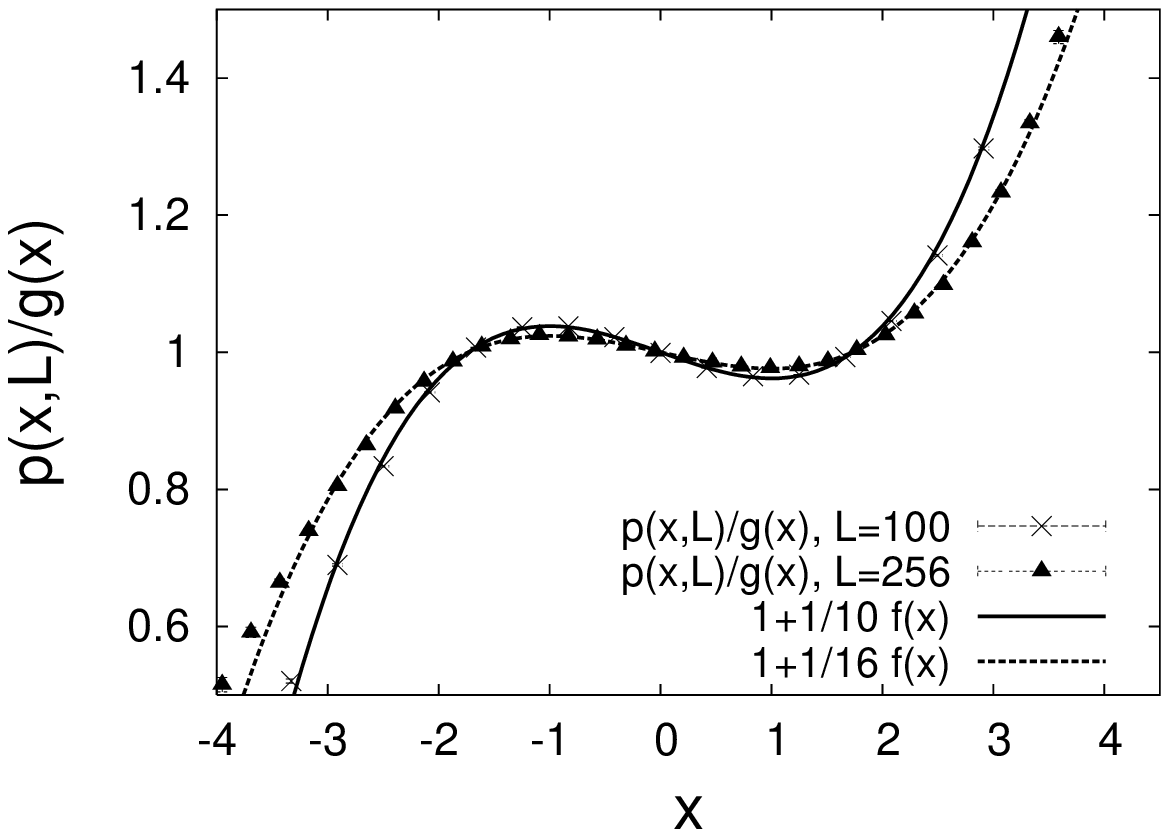}
\end{minipage}%
\caption{ (a) Results of numerical simulation of $p(x,L)$ for
$L=100, 256$: (a)  comparison of $p(x,L)$ with the Gaussian function
$g(x)$; (b) comparison of the function $p(x,L)/g(x)$ with $1 +
f(x)/\sqrt{L}$.}
\end{center}
\end{figure}

Using the approach devised in Section 2.4, we find the first
cumulant of $P(M,L)$:
\begin{equation}
\kappa_1^{2D} = \sum_j \int_0^1 \ldots \int^1_0 \prod_k dx_k
\theta(x_j-x_{j-1}) \theta(x_j - x_{j+1}) \theta(x_j - x_{j+m})
\theta(x_j - x_{j-m}) = \frac{1}{5} L
\end{equation}
where the summation over "j" and "k" extends over all lattice sites,
i.e. $j,k=1,2,3, \ldots,L$. Further on, we find that the second
cumulant is given by
\begin{eqnarray}
\kappa_2^{2D} &=& \sum_j \sum_i \int_0^1 \ldots \int^1_0 \prod_k
dx_k \theta(x_j-x_{j-1}) \theta(x_j - x_{j+1}) \theta(x_j - x_{j+m})
\theta(x_j - x_{j-m}) \times \nonumber\\
&\times& \theta(x_i-x_{i-1}) \theta(x_i - x_{i+1}) \theta(x_i -
x_{i+m}) \theta(x_i - x_{i-m})- \frac{1}{25} L^2 = \frac{13}{225} L
\end{eqnarray}
Calculation of the third cumulant is more involved \cite{hiv} and
yields
\begin{equation}
\kappa_3^{2D} = \frac{512}{32175} L.
\end{equation}
As a matter of fact, we can show that cumulants of any order grow in
proportion to the first power of $L$. Introducing next the
normalized deviation, $x=(M-\kappa_{1}^{\rm 2D})/\sqrt{\kappa_2^{2
D}}$, we seek the normalized probability distribution
$p(x,L)=P\left(\kappa_{1}^{\rm 2D}+x \sqrt{\kappa_2^{2 D}}, L
\right)$ expanding it in the Edgeworth series (cumulant expansion)
\cite{edg}. We find then that $p(x,L)$ obeys  \begin{equation}
p(x,L)\simeq g(x) \left(1+\frac{1}{\sqrt{L}}f(x)+
o\left(\frac{1}{\sqrt{L}}\right)\right), \label{cumul2}
\end{equation} where $g(x)$ is the Gaussian function $
g(x)=\exp(-x^2/2)/2\pi$ and $f(x)$ is given by
\begin{equation} f(x)=\sqrt{L} \frac{\kappa_{3}^{\rm 2D}}
{\left(\kappa_{2}^{\rm 2D} \right)^{3/2}} \frac{1}{6} (x^{3}-3 x)
=\frac{512}{32175} \left(\frac{225}{13} \right)^{3/2} \frac{1}{6}
\left(x^{3}-3 x \right). \label{cumul}
\end{equation} Note that $f(x)$ is independent of $L$ and correction
terms in Eq.(\ref{cumul2}) decay faster than $1/\sqrt{L}$.

 To check the asymptotical result in Eq.(\ref{cumul2}), we have
performed numerical simulations for the discrete 2D
permutation--generated model with periodic boundary conditions and
have computed the distribution function $p(x,L)$ numerically. In
Fig.4a we present numerical data for $p(x,L)$ and compare it against
the Gaussian function $g(x)$ for system sizes $L=100$ and $256$.
Furthermore, in Fig.4b we plot the ratio $p(x,L)/g(x)$ as the
function of $x$. One notices  that the deviation of the numerically
computed function $p(x,L)$ from the Gaussian function $g(x)$ is
indeed very small. Moreover,  the difference between the normalized
probability distribution function $p(x,L)$ and the Gaussian function
$g(x)$ is smaller the larger $L$ is.

\subsection{Gas, Liquid and Solid of Peaks}

In this last subsection we discuss a surprising collective behavior
of peaks. Consider a somewhat exotic "statistical physics" model
with the partition function:
\begin{equation}
Z = \sum_{\text{all permutations}} z^M,
\end{equation}
where $z$ is a fugacity and $M$ determines the number of peaks  for
a given distribution of natural numbers $1,2,3, \ldots, L$ on a
lattice containing $L$ sites. In one- and two-dimensions
\begin{eqnarray}
M&=& \sum_j \theta(\pi_j-\pi_{j-1}) \theta(\pi_j - \pi_{j+1}) \nonumber\\
M&=& \sum_j \theta(\pi_j-\pi_{j-1}) \theta(\pi_j - \pi_{j+1})
\theta(\pi_j - \pi_{j+m}) \theta(\pi_j - \pi_{j-m})
\end{eqnarray}
respectively.
\begin{figure}[ht]
\begin{center}
\includegraphics*[scale=0.4, angle=0]{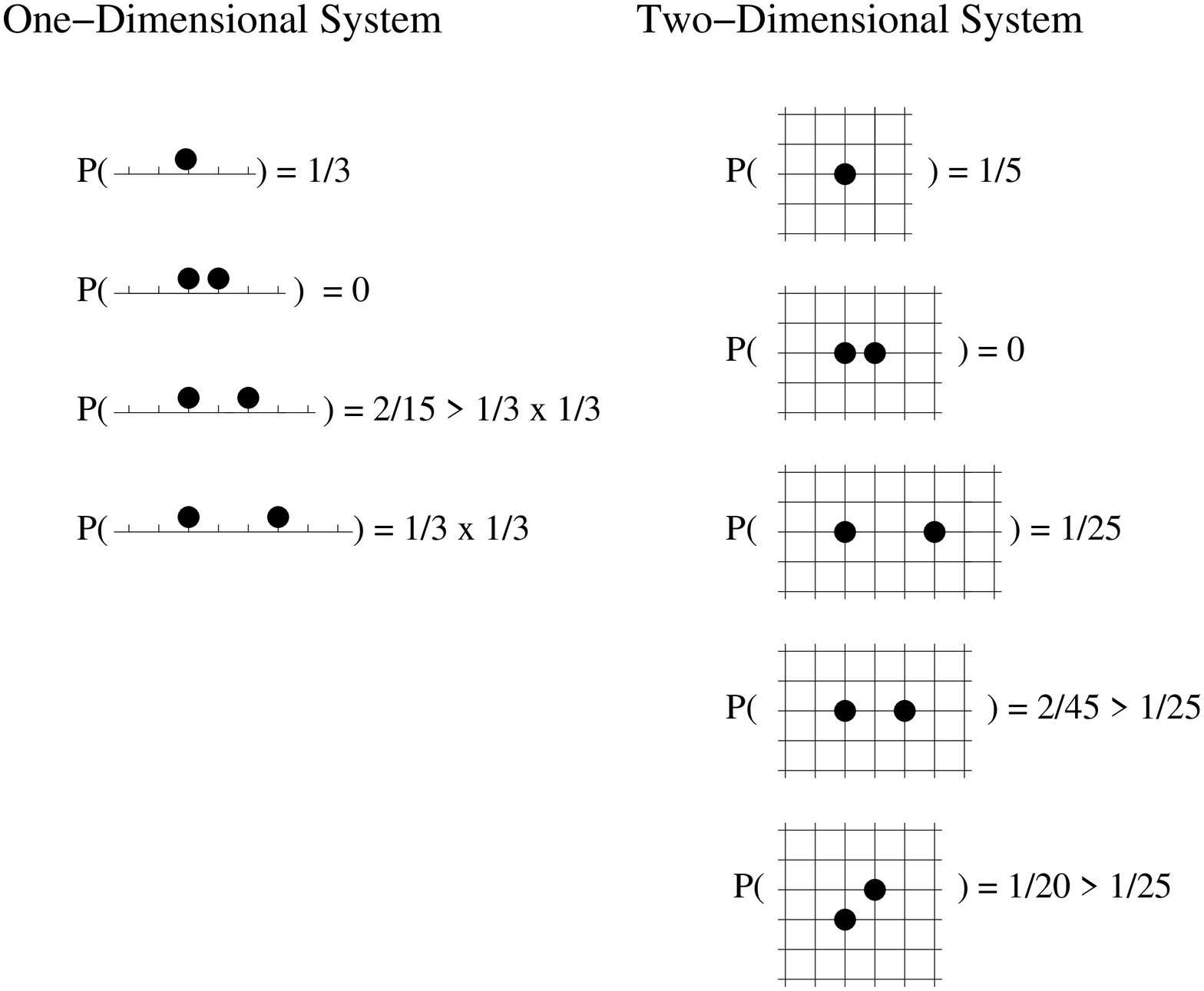}
\caption{\label{Fig5} {Probabilities of having a single peak and a
pair of isolated peaks some distance apart from each other on one-
and two-dimensional lattices. Filled circles denote peaks - sites
with numbers exceeding numbers appearing on nearest-neighboring
sites. First column presents probabilities of several configurations
in 1d, while the second one describes the probabilities of a few
possible configurations appearing in two dimensions.}}
\end{center}
\end{figure}
A striking feature of this model in two-dimensions, as communicated
to us by B.Derrida who analyzed a similar partition function
numerically, is that it undergoes  a phase transition at $z \approx
\exp(2.2)$ \cite{der}!

Below we present some arguments why such a transition may indeed
take place in two-dimensions: Note first that peaks, by definition,
can not simultaneously occupy nearest-neighboring lattice sites. It
means that if we view the peaks as some fictitious particles,  they
are not point particles occupying just a single site but rather
"hard-squares". In principle, this is already enough to produce a
phase transition, but at somewhat higher values of $z$
\cite{baxter}, since here a cost of each particle is higher
requiring a particular arrangement of numbers on five lattice sites.

Still striking, it appears that peaks experience short-range
"attractive" interactions. Note that a probability of having a
single isolated peak is equal to $1/5$. The probability of having
two peaks separated by distance exceeding $l=2$ is $1/25 = 1/5^2$,
which means that at such distance the peak do not feel each other.
On the other hand, the probability of having two peaks at
next-nearest-neighboring sites or in corners of each lattice cell,
is greater than $1/25$, which means that peaks effectively "attract"
each other! In Fig.4 we summarize the results for the probabilities
of several simple configurations of peaks.

This "attractive" force between peaks can be thought off as a sort
of a depletion force acting between colloidal particles: by
definition, each peak is a site with a number which is greater than
numbers on four neighboring sites. When two peaks are
next-nearest-neighboring sites, they have a "common" site with a
number which is restricted to be less than the least of the numbers
on the peak sites, i.e. may attain less values than numbers on the
sites around an isolated peak. This effect is even more pronounced
for peaks occupying sites in the corners of a lattice cell - here
these two peaks have two "common" numbers and each of them has to be
less than the least of them.

Note finally that for clusters containing several peaks, the
presence of such common sites makes the interactions between peaks
non-additive and dependent on the specific features of the cluster's
geometry. Thus it is not evident at all whether the phase transition
is in the Ising universality class.

\section{Acknowledgments}

The authors wish to thank B.Derrida, P.H\"anggi, P.Krapivsky and
Z.Racz for helpful discussions and remarks. The work is partially
supported by the grant ACI-NIM-2004-243 "Nouvelles Interfaces des
Math\'ematiques" (France).

\end{document}